\documentstyle[epsfig,rotating,a4,12pt]{article}

\begin{document}
\title{A Proposal for a Different Chi-square Function for Poisson Distributions}
\author{F. M. L. Almeida Jr.\footnote{email: marroqui@if.ufrj.br}
 and M. Barbi\footnote{Now at McGill University, Canada. email:
 barbi@mail.desy.de}\\
Instituto de F\'\i sica - Universidade Federal do Rio de Janeiro\\
Ilha do Fund\~ao,Rio de Janeiro,\\
21945-970, RJ, BRAZIL\\
\\
M. A. B. do Vale\footnote{email: aline@lafex.cbpf.br}\\
Centro Brasileiro de Pesquisas F\'\i sicas\\
 Rua Xavier Sigaud 150, Urca, Rio de Janeiro\\
22290-180, RJ, BRAZIL}

\date{ }
\maketitle

\begin{abstract}
We obtain an approximate Gaussian distribution from a Poisson distribution
after doing a change of variable.  A new chi-square function is obtained
which can be used for parameter estimations and goodness-of-fit testing
when adjusting curves to histograms.  Since the new distribution is 
approximately Gaussian we can use it even when the bin contents are small.
The corresponding chi-square function can be used for curve fitting.
This chi-square  function is simple to implement and presents a fast convergence
 of the parameters to the correct value, especially for the parameters
 associated with the width of the
fitted curve.  We present a Monte Carlo comparative study of the 
fitting method introduced here and two other methods for three types of curves:
Gaussian, Breit-Wigner and Moyal, when each bin content obeys a Poisson
distribution. It is also shown that the new method and the other two  
converge to the same result when the number of events increases.
\end{abstract}

\pagebreak
\section{Introduction}
~
Nowadays there is an intensive experimental effort to search for new
phenomena in many fields of particle physics.  These
signals should be rare events where one needs to treat low statistics
data and  non-Gaussian errors.  In order to handle these data one needs
special care, since all the usual Gaussian procedures and techniques are
no longer valid.  On the other hand, it turns out that the minimization 
of the chi-square functions is normally simple, 
relatively fast and a very familiar way of fitting
data, estimating parameters and their errors so as the goodness-of-fit
testing. The $\chi^2$ of fitting a theoretical
curve to experimental data in order to estimate some parameter values
is to minimize the function

\begin{equation}
\chi^2_G = \sum_{i=1}^{N}
\frac{(f(x_{i},\vec{\alpha})-n_i)^2}{\sigma_i^2}
\label{1x1}
\end{equation}

\noindent where $f(x_i,\vec{\alpha})$ is a known function(model 
prediction) calculated at
the point $x_i$ and $\vec{\alpha}$ is a vector of the parameters one
wants to obtain; $n_i$ is the measured experimental value associated with
the bin located at $x_i$ and $N$ is the number of bins in the range of interest.
This method has been largely used but presents some limitations such as
the assumption that each bin content must obey a Gaussian distribution of spread
$\sigma_i$ or the measurement errors are at least almost ''Gaussian''. If
the contents of each bin obeys a Poisson distribution some authors
\cite{Brandt} recommend that each bin content must have at least a
statistically significant number of entries $n_i$  such as that $\sigma_i
\approx \sqrt{n_i}$, where the asymptotically limiting case for large
number of measurements starts, and the square root of the variance can be
considered as a good interval estimation. For values smaller than these,
there are suggestions that one must use bins of variable sizes such that
each bin content be greater than the statistically significant number of
entries $n_i$. The disadvantage of these suggestions is that one can
lose important information about the structure of the studied
distribution. There is no rule of thumb for the bin width or for the
ideal number of bins in which the region of interest should be divided. There is
yet another hint that the ideal case is to use bins of variable width of
equal probability contents \cite{Eadie}.\par

In a very interesting paper, Baker and Cousins\cite{Baker}, 
 call attention and discuss some topics such as
point estimation, confidence interval estimation, 
goodness-of-fit testing, biased estimation, etc. 
when fitting curves to histograms using chi-square function.\par

They presented a $\chi^2_{BC}$ function for fitting histograms when
the bin contents obey a Poisson distribution. They defined a Poisson likelihood 
chi-square which is given by Eq.(\ref{1x2}) below,

\begin{equation}
\chi^2_{BC}=2 \sum_{i=1}^N\left[f(x_i,\vec\alpha)-n_i+n_i 
log\left(n_i/f(x_i,\vec\alpha)\right)\right]
\label{1x2}
\end{equation}

This function behaves asymptotically as a chi-square distribution and
 then can be used for estimation and goodness-of-fit testing.\\

The chi-function presented in this paper has also asymptotically a behavior
like the classical chi-square function, a fast convergence to the correct value,
much less fluctuations when compared with $\chi^2_G$ and it works 
also when one has distributions with long tails where bin contents can
be very low.  It is of easy implementation in any minimization program.\\

After this introduction, we demonstrate in section  2 how one can
transform a non-Gaussian pdf into an approximate Gaussian pdf. 
Section 3 is devoted to obtain the chi-square function
for the approximate Gaussian pdf from a Poisson distribution 
and discuss some of its
characteristics. The next section, 4, we compare with the results obtained
using the Eq.(\ref{1x1}) and Eq.(\ref{1x2}), 
and the new chi-square function for different number of
entries in the histograms.  This comparison is made using Monte Carlo events generated
according to some distributions with known parameters, 
as suggested by \cite{Baker} and in section 5 we present the conclusions.
 In the appendix we obtain the equivalent chi-square expression
 for a binomial distribution.

\section{Obtaining an approximate Gaussian distribution}
~
The basic motivation is to transform, via a variable transformation,
a non-Gaussian probability density
function(pdf) into an approximately Gaussian pdf preserving the
probability even when one has small number of events \cite{Eadie,Box}. Let us
consider a non-Gaussian pdf $p(x)$ and one wants to obtain a
transformation such that the probability is preserved Eq.(\ref{2x1}) and
that the new pdf $q(z)$ should be approximately Gaussian:

\begin{equation}
q(z)dz = p(x)dx
\label{2x1}
\end{equation}
Then $q(z)$ can be written as
\begin{equation}
q(z) = p(x) \left\arrowvert\frac{dx}{dz}\right\arrowvert
\label{2x2}
\end{equation}

When a pdf is unimodal and obeys some regularity conditions \cite{Box},
the logarithm of it is
approximately quadratic so that

\begin{equation}
log(p(x)) \approx log(p(\hat{x}))-
\frac{1}{2}\left (-\frac{\partial^2 log(p(x))}{\partial x^2 }\right
)_{\hat x}(x-\hat{x})^2
\label{2x3}
\end{equation}

\noindent where $\hat{x}$ is the point associated to the maximum
 of $p(x)$ and one can define the following  quantity

\begin{equation}
J(\hat{x}) = \left (-\frac{\partial^2 log(p(x))}{\partial x^2}\right
)_{\hat{x}}
\label{2x4}
\end{equation}

On the other hand, the logarithm of a Normal pdf $g(x)$, with mean $\hat\mu$
and standard deviation $\sigma$, is of the form:

\begin{equation}
log(g(x)) = const - \frac{1}{2 \sigma^2} (x-\hat{\mu})^2
\label{2x5}
\end{equation}

\noindent so that  given the location parameter $\hat\mu$,
 it is completely determined by its standard deviation $\sigma$.

A comparison between  Eq.(\ref{2x3}) and Eq.(\ref{2x5}) shows that the
variance of the pdf $p(x)$ is approximately equal to $J^{-1}(\hat{x})$.
Let us suppose now that $z(x)$ is a one-to-one transformation between $x$
and $z$, then using the chain rule for derivatives one gets the relation

\begin{eqnarray}
J(\hat{z}) & = & \left ( -\frac{\partial^2 log(p(x))}{\partial x^2}\right
)_{\hat x} \left\arrowvert \frac {dx}{dz}\right\arrowvert^2_{\hat z}\cr
 & = & J(\hat x)\left\arrowvert \frac {dx}{dz}\right\arrowvert^2_{\hat z}
\label{2x6}
\end{eqnarray}

Let us choose $z(x)$ such that

\begin{equation}
\left (\frac{dx}{dz}\right )_{\hat z} = J^{-1/2}(\hat{x})
\label{2x7}
\end{equation}

This choice is made so as to make $J(\hat z)$ independent of $\hat z$,
the standard deviation equal to one and the new distribution $q(z)$ 
approximately translation invariant along the $z$ axis. Thus the metric
can be obtained from the relationship obeying the above conditions

\begin{equation}
\frac{dz}{dx} = J^{1/2}(x)
\label{2x8}
\end {equation}

\begin{equation}
z = \int^x  J^{1/2}(t) dt
\label{2x9}
\end{equation}

\section{The Chi-square Function}
As an example, let us apply the above prescription to a Poisson pdf 
 given by

\begin{equation}
p_P(x) = \frac{x^k e^{-x}}{\Gamma(k+1)}
\label{2x10}
\end{equation}

\noindent which means that after observing $k$ events one has a pdf of
the estimated mean parameter $x$.  Now one wants to find a Gaussian like
pdf through a transformation of $x$. The location of the $p_P(x)$ maximum
is easily shown to be at $\hat x= k$ and the term associated to the
second derivative is

\begin{equation}
-\frac{\partial^2 log(p(x))}{\partial x^2} = \frac{k}{x^2}
\label{2x11}
\end{equation}

 Using the fact that $\hat{x} = k$, one obtains

\begin{equation}
J(\hat{x}) = \frac{1}{k}
\label{2x12}
\end{equation}

\noindent then

\begin{equation}
J^{1/2}(x) = \frac{1}{\sqrt{x}}
\label{2x13}
\end{equation}

Using Eq(\ref{2x7}), the one-to-one transformation is obtained as

\begin{equation}
\frac{dz}{dx} = \frac{1}{\sqrt{x}}
\label{2x14}
\end{equation}

\begin{equation}
z = \int^x \frac{1}{\sqrt{t}}dt
\label{2x15}
\end{equation}

\begin{equation}
z = 2 \sqrt{x}
\label{2x16}
\end{equation}

\noindent and the inverse transformation is

\begin{equation}
x = \left (z \over 2\right )^2
\label{2x17}
\end{equation}

Reusing the above expression in Eq.({\ref{2x10}) and using
Eq.(\ref{2x2}), one obtains an approximately Gaussian expression $q_P(z)$
associated to the Poisson pdf $p_P(x)$, which is

\begin{equation}
q_P(z) = \frac{\left(z/2\right)^{2k+1} e^{-\left (z/2\right
)^2}}{\Gamma(k+1)}
\label{2x18}
\end{equation}

It is not difficult to show that the above expression is normalized,
$J(\hat z)=1$ and is translation invariant by construction. The
approximate Gaussian and exact Gaussian distributions,$q_P(z)$ and
$g(z)$, respectively, are shown in Fig. 1 for different values of $k$ ($k
= 0,1,2,5$ and $10$). The worst approximation occurs at $k=0$ but it gets
better very fast as $k$ increases.

This pdf $q_P(z)$ has a maximum at $\hat{z} = \sqrt{4k+2}$ which 
corresponds to $x_{\hat z} = k+ 1/2$. It is interesting to note that
$x_{\hat z}$ is between $\hat x=k$ and the median of the Poisson pdf,
$x_m \approx k + 2/3$, i.e., the maximum of $p_P(x)$ is not directly
related to the maximum of $q_P(z)$ via Eq.(\ref{2x16}) and
Eq.(\ref{2x17}).

It is not difficult either to obtain an analytical expression for the
confidence intervals for roughly $68.3\%$ of confidence level since the
obtained $q_P(z)$ has a maximum at $\hat{z} = \sqrt{4k+2}$ and a standard
deviation equal to unit, one gets

\begin{equation}
\left [z_{min},z_{max}\right ] =\left [ \sqrt{4k+2}-1,\sqrt{4k+2}+1\right
]
\label{2x19}
\end{equation}
Taking the inverse transformation one gets the corresponding interval
associated with the original Poisson pdf

\begin{equation}
\left[x_{min},x_{max}\right ]=
\left[(z_{min}/2)^2,(z_{max}/2)^2\right ]
\label{2x20}
\end{equation}

The confidence level calculated according to Eq.(\ref{2x19})and
Eq.(\ref{2x20}) is shown in Table 1 for different values of $k$, and so are
the probability contents of the calculated interval. One can see that
these intervals overestimate the $68.27\%$ confidence level but converge
to it as $k$ increases.

Let us suppose that one wants to fit a set of data when the contents of
each bin obeys a Poisson distribution. If the contents of each bin has
small numbers of events, one can no longer use as its standard deviation
$\sigma_i = \sqrt{n_i}$, since the errors are asymmetrical and
consequently one can not use the least-square fit method, Eq.(\ref{1x1}), either since it
works only for Gaussian pdf. If one insists one could get large
deviations for the estimated parameters as is shown in the
figures of the next section. After taking the above
transformation, the contents of each bin is approximately Gaussian.

Using the likelihood ratio test theorem \cite{Eadie,Baker}, one gets 
the following expression in terms of bin contents $n_i$
\begin{equation}
\chi^2_P=-2 \sum_{i=1}^N log(\lambda_i)
\end{equation}
\noindent where 
\begin{equation}
\lambda_i=\frac{\left({\sqrt{4f(x_i,\vec\alpha)+2}\over 2}\right)^{2n_i+1}
e^{-\left({\sqrt{4f(x_i,\vec\alpha)+2}\over 2}\right)^2}}
{\left({\sqrt{4n_i+2}\over 2}\right)^{2n_i+1}
e^{-\left({\sqrt{4n_i+2}\over 2}\right)^2}}
\end{equation}

\noindent which gives

\begin{equation}
\chi^2_P=\sum_{i=1}^N\left[2(f(x_i,\vec\alpha)-n_i)+
(2n_i+1)log\left({2n_i+1\over2 f(x_i,\vec\alpha)+1}\right)\right]
\label{2x21}
\end{equation}

\noindent which asymptotically behaves like a chi-square distribution.
This expression is similar to Eq.(\ref{1x2}) obtained
by \cite{Baker}.

One can also derive an equivalent expression for 
a binomial distribution.  This is shown in the appendix.

\section{Comparing the different chi-square functions}
~
Let us now compare the results obtained by minimizing Eq.(\ref{1x1}),
Eq.(\ref{1x2}) and Eq.(\ref{2x21}) when the bin contents
obeys a Poisson distribution.
This comparison was made for three different
curves: Gaussian, Breit-Wigner and
Moyal, Eq.(\ref{3x1}),Eq.(\ref{3x2}) and
Eq.(\ref{3x3}), respectively.
\begin{equation}
f_{G}(x) \propto e^{\displaystyle{-\frac{(x-\mu_G)^2}{2 \sigma_G^2}}}
\label{3x1}
\end{equation}

\begin{equation}
f_{BW}(x) \propto \frac{1} {(2 \sigma_{BW}^{2}+(x-\mu_{BW})^ 2) }
\label{3x2}
\end{equation}

\begin{equation}
f_{M}(x) \propto e^{\displaystyle{{-h-e^{-h}}}}
\label{3x3}
\end{equation}

\noindent where $h=(x-\mu_{M})/\sigma_{M}$.\\

The first two are symmetrical curves with ''short'' and long
''tails'', respectively, while the last one is an asymmetrical function
with a left ''short'' and a right ''long'' tails.

The parameters $\{ \mu_j \}$ are associated to the maximum value of the
distribution and $\{ \sigma_j \}$ are related to the spread of
the distribution, where $j=G,BW,M$.

One generates random points with known $\bar \mu_j$ and $\bar \sigma_j$
according to each of the above distributions and filling  histograms with
100 bins, see Table 2.

The number of entries ranges from 20 to $10^4$ and for each fixed number
of entries, $10^4$ sets of points were generated.

For each fixed number of entries we calculate the average value of the
fitted parameters $\mu_j$ and $\sigma_j$ which are the estimators of $\bar
\mu_j$ and $\bar \sigma_j$, respectively, using Eq.(\ref{1x1}), Eq.(\ref{1x2})
and Eq.(\ref{2x21}) and the MERLIN optimization package \cite{merlin}.

The fit was done from the first bin to the last bin content different from zero,
although there could exist bins of contents equal to zero in between, 
except for Eq.(\ref{1x1}) where the bins of contents equal to zero
were excluded in order to avoid singularities. 
One also calculates the expected mean errors defined as  $\Delta
\sigma_j=\sqrt{<\sigma_j-\bar\sigma_j>^2}$ and $\Delta
\mu_j=\sqrt{<\mu_j-\bar\mu_j>^2}$ of the fitted  $\mu_j$ and $\sigma_j$
with respect to the ''true''  known $\bar \mu_j$ and $\bar \sigma_j$,
fixing the number of entries.

These results are summarized in Figs.2-13 in terms of the number of
entries. We can see clearly that the minimization of $\chi^2_P$
shows faster or equal convergence to ''true'' value  
and smaller or equal expected mean errors as the number of entries
increase than when using $\chi^2_{BC}$. Both these functions
are systematically much better than $\chi_G$
for the convergence and expected mean errors of the parameters. 
For small number of entries, we can also notice
that the parameters $\sigma_i$ are systematically overestimated for
all the shown cases while $\mu_M$ for the Moyal case is underestimated.
All the chi-square functions here presented converge systematically to
the correct value as the number of entries increase.
One can observe in all figures that all three methods coincide as the
number of entries increase. We can clearly see the advantage of using
the chi-square function introduced here. It converges equally well or faster and  has 
equal or smaller expected mean errors than the other two methods.
 Besides, Eq.(\ref{2x21}) is also of easy
implementation in optimization programs .

\section{Conclusions}
This article presented an improvement over the usual minimization of chi-square 
function technique for
fitting functions in order to extend its applicability to low statistics
data when one has asymmetrical errors.  This new method is obtained
through the change of variables such that one gets approximately Gaussian
pdf when transforming the original pdf.  The approximate Gaussian pdf is
obtained associated to a Poisson pdf and a chi-square function
 is adapted for this new
Gaussian like pdf.  Monte Carlo generated
events show an improvement in low statistics region, as expected,
although the results converge to  the ``true'' value as the number of events 
increases.  This method has shown a fast convergence to the correct
parameter values especially to the parameter associated to the curve
spread, $\sigma_j$. The proposed chi-square function is consistent since the
fitted parameters converge to the true value of the parameters 
and their expected mean errors decreases as the number of observations increases.  
The results are simple and of easy use in standard
optimization procedures.  A similar result was obtained in the appendix
for  binomial distributions.
\\

\textit{Acknowledgments:} The authors are very grateful for discussions with
A. Ramalho and A. Vaidya.
  This work was supported in part by the following Brazilian
agencies: CNPq, FINEP, FAPERJ, FJUB and CAPES.

\newpage
\appendix
\section{Chi-square Function for Binomial Distribution}
Let us now derive a $\chi^2_B$ for the case when each bin content obeys a
binomial distribution, this is the usual case when dealing with
uncorrelated normalized data. The binomial distribution $p_B(x)$, normalized with
respect to $x$, can be written as
\begin{equation}
p_B(x)=\frac{\Gamma(n+2)x^k(1-x)^{n-k}}{\Gamma(k+1)\Gamma(n-k+1)}
\label{A.1}
\end{equation}
\noindent where $x\in[0,1]$, then

\begin{eqnarray}
J(\hat{x}) &=& \left (-\frac{\partial^2 log(p_B(x))}{\partial x^2}\right
)_{\hat{x}}\cr
&=& \frac{1}{\hat x(1-\hat x)}
\label{A.2}
\end{eqnarray}

Supposing again that $z(x)$ is a
one-to-one transformation, then

\begin{eqnarray}
J(\hat{z}) &=& -\left (\frac{\partial^2 log(p_B(x))}{\partial x^2}\right
)_{\hat x} \left\arrowvert \frac {dx}{dz}\right\arrowvert^2_{\hat x}
\label{A.3}
\end{eqnarray}

Let us choose $dx/dz$ such that

\begin{eqnarray}
\frac{dx}{dz} &=& J^{-1/2}({x})\cr
&=& \sqrt{x(1-x)}
\label{A.4}
\end{eqnarray}

\noindent then inverting the above equation one gets

\begin{equation}
\frac{dz}{dx} = J^{1/2}(x)
\label{A.5}
\end {equation}

Integrating the latter expression
\begin{eqnarray}
z(x) &=& \int^x  \frac{1}{\sqrt{t(1-t)}} dt\cr
     &=& \arcsin(2x-1)
\label{A.6}
\end{eqnarray}

Then the approximately Gaussian pdf associated to the binomial pdf above
is

\begin{equation}
q_B(z)=
\frac{\Gamma(n+2)}{2^{n+1}\Gamma(k+1)\Gamma(n-k+1)}(1+\sin(z))^k(1-\sin(z
))^{(n-k)}\cos(z)
\label{A.7}
\end{equation}
where $z\in[-\pi/2,\pi/2]$.
This function has a variance $\sigma^2=1/(n+1)$ which is independent
of $k$ and its maximum is located at

\begin{eqnarray}
\hat z&=&\arcsin\left (\frac{2k-n}{n+1}\right )\cr
&=&  \arcsin\left (\frac{2k/n-1}{1+1/n}\right )
\label{A.8}
\end{eqnarray}

The approximate 68.27\% confidence limits are given by the analytical
expression
\begin{equation}
\arcsin\left(\frac{2k/n-1}{1+1/n}\right)\mp\frac{1}{\sqrt{n+1}}
\end{equation}

\noindent and after using the likelihood ratio test theorem, one obtains the
 chi-square function for a binomial pdf

\begin{eqnarray}
\chi^2_B&=&\sum_{i=1}^N\left[(2n_i+1)log\left({2n_i+1\over 2nf(x_i,\vec\alpha)+1}\right)\right.\\
&+&\left.\left(2(n-n_i)+1\right)log\left({2(n-n_i)+1\over 2n\left(1-f(x_i,\vec\alpha)\right)+1}\right)\right]
\end{eqnarray}

\noindent where $0 \leq f(x_i,\vec\alpha) \leq 1$ are the estimated parameters.

\vfill
\newpage
\section*{Figure captions}
\newcounter{ijk}
\begin{list}%
{Fig.\arabic{ijk}\
}{\usecounter{ijk}\setlength{\rightmargin}{\leftmargin}}

\item The approximate Gaussian probability density function(pdf)
associated to a Poisson pdf for $k=0,1,2,5$ and $10$ and the exact
Gaussian distribution with variance equal to one.

\item The average value of the fitted Gaussian parameter $\mu_G$ as a
function of the number of entries. The curves P, BC and
LSM correspond to chi-square functions $\chi^2_P$, $\chi^2_{BC}$ and
$\chi^2_G$, respectively.

\item The expected mean error $\Delta\mu_G$ of the fitted parameter
$\mu_G$ as a function of the number of entries. The curves P, BC and
LSM correspond to chi-square functions $\chi^2_P$, $\chi^2_{BC}$ and
$\chi^2_G$, respectively.

\item Same as Fig.2 but for the parameter $\sigma_G$.

\item Same as Fig.3 but for the parameter $\sigma_G$.

\item The average value of the fitted Breit-Wigner parameter $\mu_{BW}$
as a function of the number of entries.  The curves P, BC and
LSM correspond to chi-square functions $\chi^2_P$, $\chi^2_{BC}$ and
$\chi^2_G$, respectively.

\item The expected mean error $\Delta\mu_{BW}$ of the fitted parameter
$\mu_{BW}$ as a function of the number of entries.  The curves P, BC and
LSM correspond to chi-square functions $\chi^2_P$, $\chi^2_{BC}$ and
$\chi^2_G$, respectively.

\item Same as Fig.6 but for the parameter $\sigma_{BW}$.

\item Same as Fig.7 but for the parameter $\sigma_{BW}$.

\item The average value of the fitted Moyal parameter $\mu_{M}$ as a
function of the number of entries.  The curves P, BC and
LSM correspond to chi-square functions $\chi^2_P$, $\chi^2_{BC}$ and
$\chi^2_G$, respectively.

\item The expected mean error $\Delta\mu_{M}$ of the fitted parameter
$\mu_{M}$ as a function
of the number of entries.  The curves P, BC and
LSM correspond to chi-square functions $\chi^2_P$, $\chi^2_{BC}$ and
$\chi^2_G$, respectively.

\item Same as Fig.10 but for the parameter $\sigma_{M}$.

\item Same as Fig.11 but for the parameter $\sigma_{M}$.

\end{list}

\newpage

\begin{table}
\begin{center}
\begin{tabular}{||c||c|c|c|c|c||} \hline
\ & & & & & \\
\ & $z_{min}=$ & $z_{max}=$ & $x_{min}=$ & $x_{max}=$ &
$\int_{y_{min}}^{y_{max}}q_P(z)dz$\\
$k$ &  &  & &  & \\
\ & $\sqrt{4k+2}-1$ & $\sqrt{4k+2}+1$ & $(z_{min}/2)^2$
 & $(z_{max}/2)^2$ & $\int_{x_{min}}^{x_{max}}p_P(x)dx$ \\
\ & & & & &\\
\hline
\hline
0& 0.414213562& 2.414213562& 0.04289321873& 1.457106781&0.7251045244\\
\hline
1& 1.449489743& 3.449489743& 0.5252551288& 2.974744873&0.6990903884\\
\hline
2& 2.162277660& 4.162277660& 1.168861170& 4.331138830&0.6926877287\\
\hline
3& 2.741657387& 4.741657387& 1.879171307& 5.620828695&0.6898630257\\
\hline
4& 3.242640687& 5.242640687& 2.628679658& 6.871320343&0.6882789039\\
\hline
5& 3.690415760& 5.690415760& 3.404792120& 8.095207880&0.6872666311\\
\hline
6& 4.099019514& 6.099019514& 4.200490245& 9.299509758&0.6865643067\\
\hline
7& 4.477225575& 6.477225575& 5.011387213& 10.48861279&0.6860486195\\
\hline
8& 4.830951895& 6.830951895& 5.834524053& 11.66547595&0.6856539615\\
\hline
9& 5.164414003& 7.164414003& 6.667792998& 12.83220700& 0.6853422256\\
\hline
10& 5.480740698& 7.480740698& 7.509629650& 13.99037035&0.6850897795\\
\hline
20& 8.055385138& 10.05538514& 16.22230743& 25.27769258&0.6839191522\\
\hline
30& 10.04536102& 12.04536102& 25.22731950& 36.27268053&0.6835159909\\
\hline
40& 11.72792206& 13.72792206& 34.38603895& 47.11396103&0.6833119043\\
\hline
50& 13.21267040& 15.21267040& 43.64366478& 57.85633518&0.6831886429\\
\hline
60& 14.55634919& 16.55634919& 52.97182543& 68.52817463&0.6831061362\\
\hline
70& 15.79285562& 17.79285562& 62.35357215& 79.14642778&0.6830470354\\
\hline
80& 16.94435844& 18.94435844& 71.77782073& 89.72217918&0.6830026194\\
\hline
90& 18.02629759& 20.02629759& 81.23685120& 100.2631488&0.6829680213\\
\hline
100& 19.04993766& 21.04993766& 90.72503123& 110.7749689&0.6829403087\\
\hline
\end{tabular}
\caption{Approximate 68.3\% confidence limits, $[z_{min},z_{max}]$ and
$[x_{min},x_{max}]$, for the almost  Gaussian and Poisson distributions,
respectively, for different $k$ values.}
\label{table:pclimits}
\end{center}
\end{table}
\newpage
\begin{table}
\begin{center}
\begin{tabular}{|c|c|c|c|} \hline
\ & & & \\
Distribution & $\bar\mu_j$ & $\bar\sigma_j$ & Range\\
\ & & & \\
\hline
\hline
\ & & & \\
Gauss& $0.0$ & $1.0$ & $[-5.0,5.0]$\\
\ & & & \\
\hline
\ & & & \\
Breit-Wigner& $50.0$ & $2.0$ & $[0.0,100.0]$\\
\ & & & \\
\hline
\ & & & \\
Moyal& $2.0$& $0.5$ & $[0.0,8.0]$\\
\ & & & \\
\hline
\end{tabular}
\caption{Distributions, their parameters and the range divided in 100
bins
according to which the random numbers were generated.}
\label{table:parameters}
\end{center}
\end{table}
\vfill

\begin{figure} 
\begin{center}
\begin{sideways}
\mbox{\epsfig{file=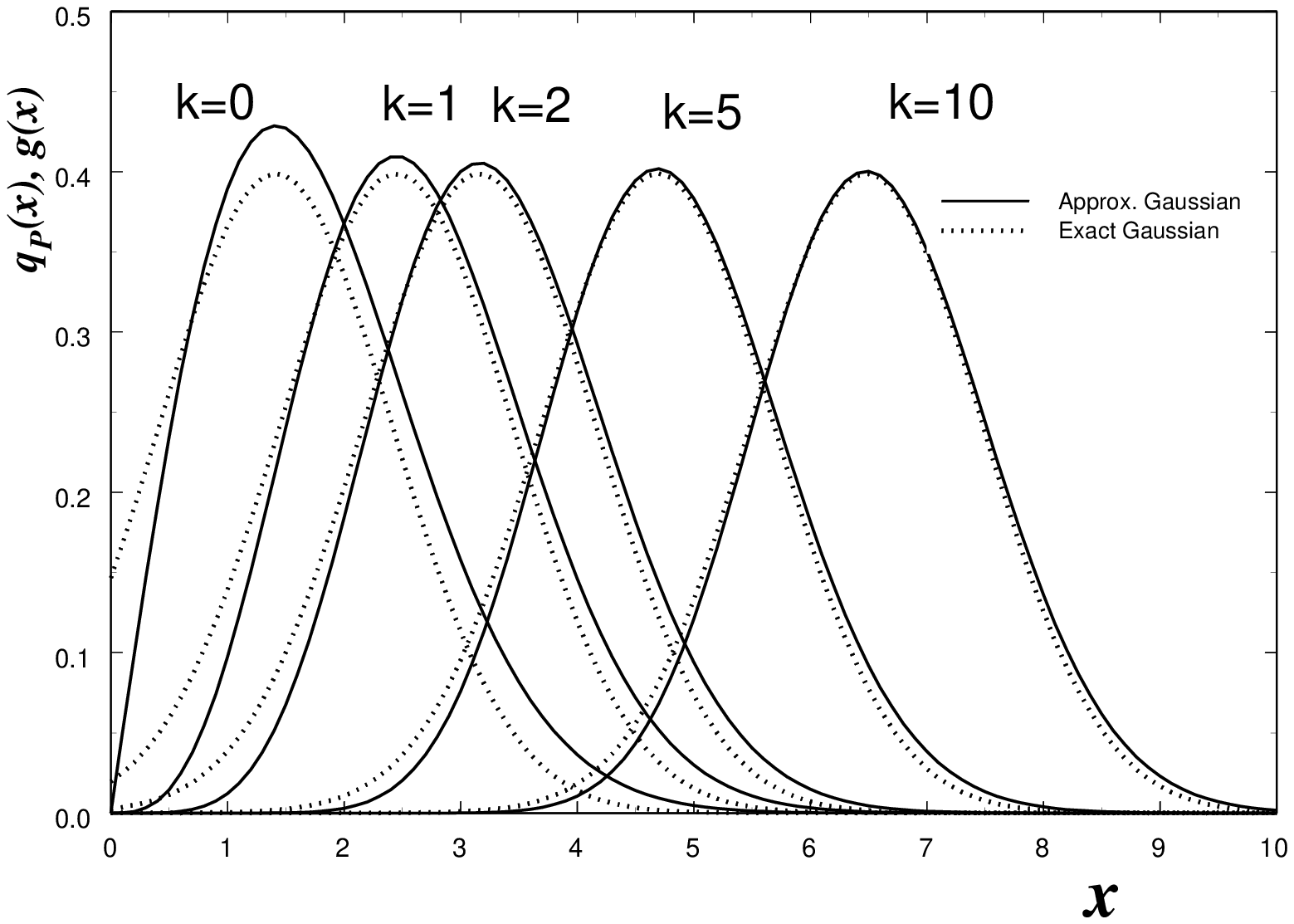}} 
\end{sideways}
\end{center}
\caption{ }
\end{figure}

\begin{figure} 
\begin{center}
\begin{sideways}
\mbox{\epsfig{file=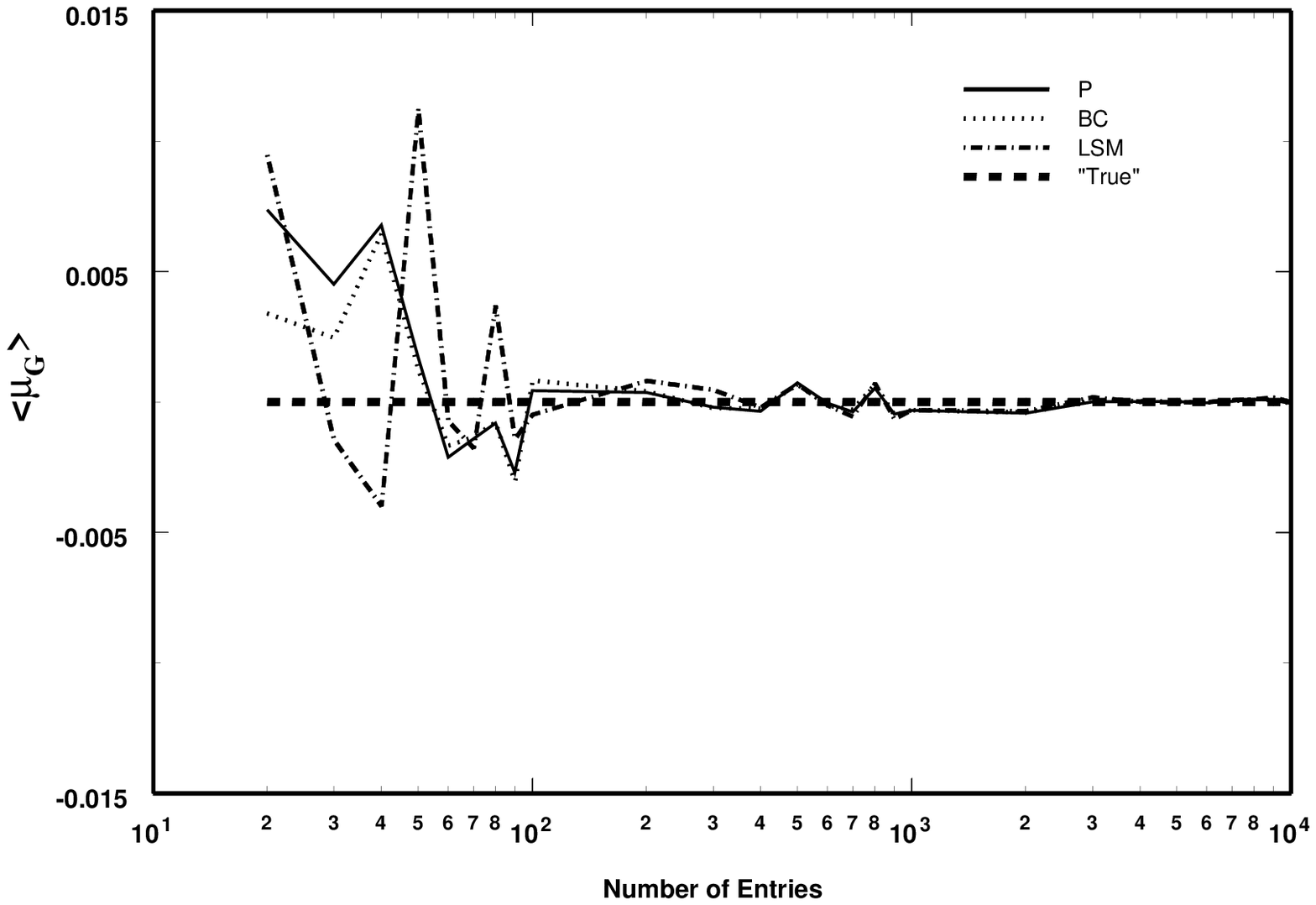}} 
\end{sideways}
\end{center}
\caption{ }
\end{figure}

\begin{figure} 
\begin{center}
\begin{sideways}
\mbox{\epsfig{file=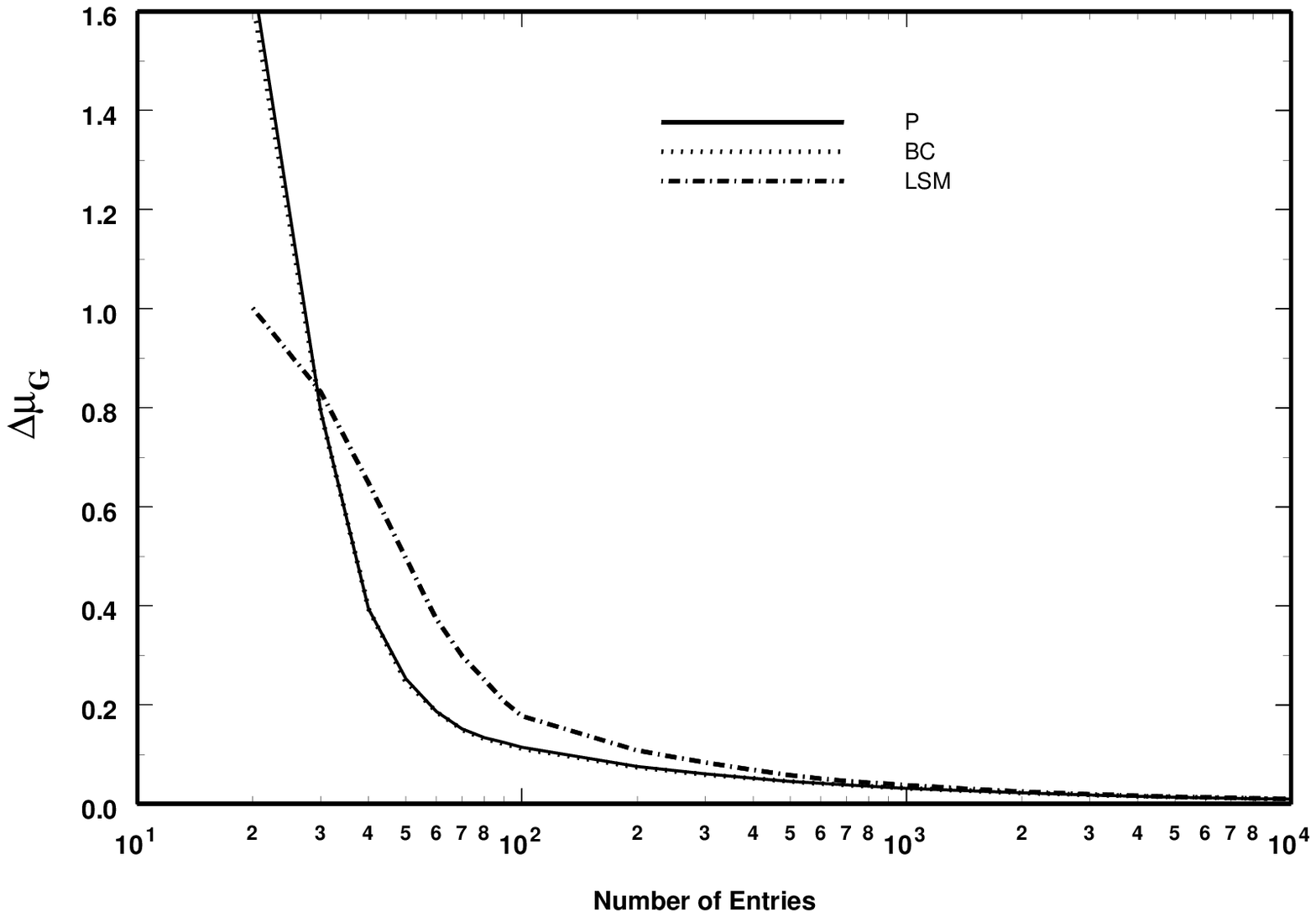}} 
\end{sideways}
\end{center}
\caption{ }
\end{figure}

\begin{figure} 
\begin{center}
\begin{sideways}
\mbox{\epsfig{file=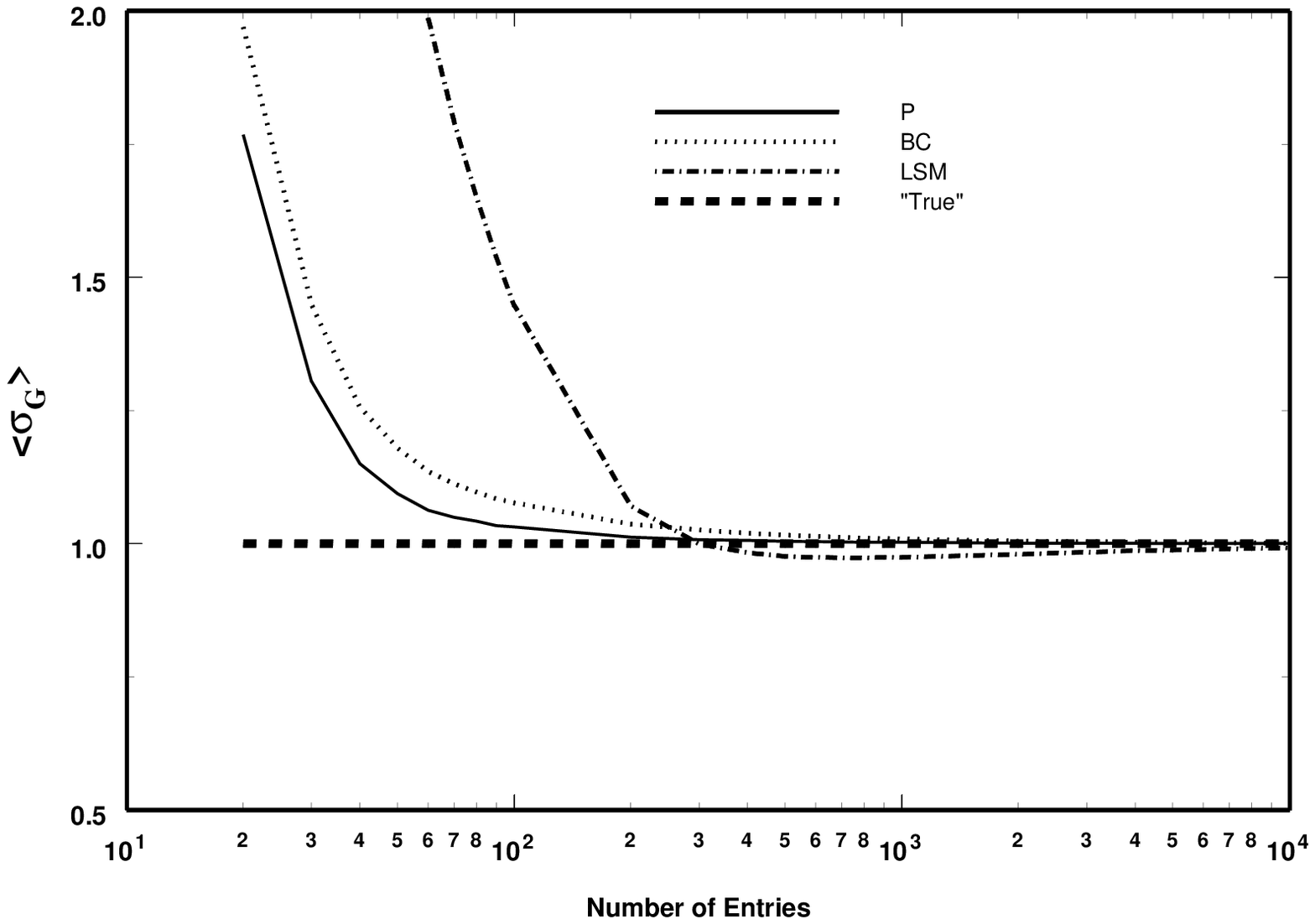}} 
\end{sideways}
\end{center}
\caption{ }
\end{figure}

\begin{figure} 
\begin{center}
\begin{sideways}
\mbox{\epsfig{file=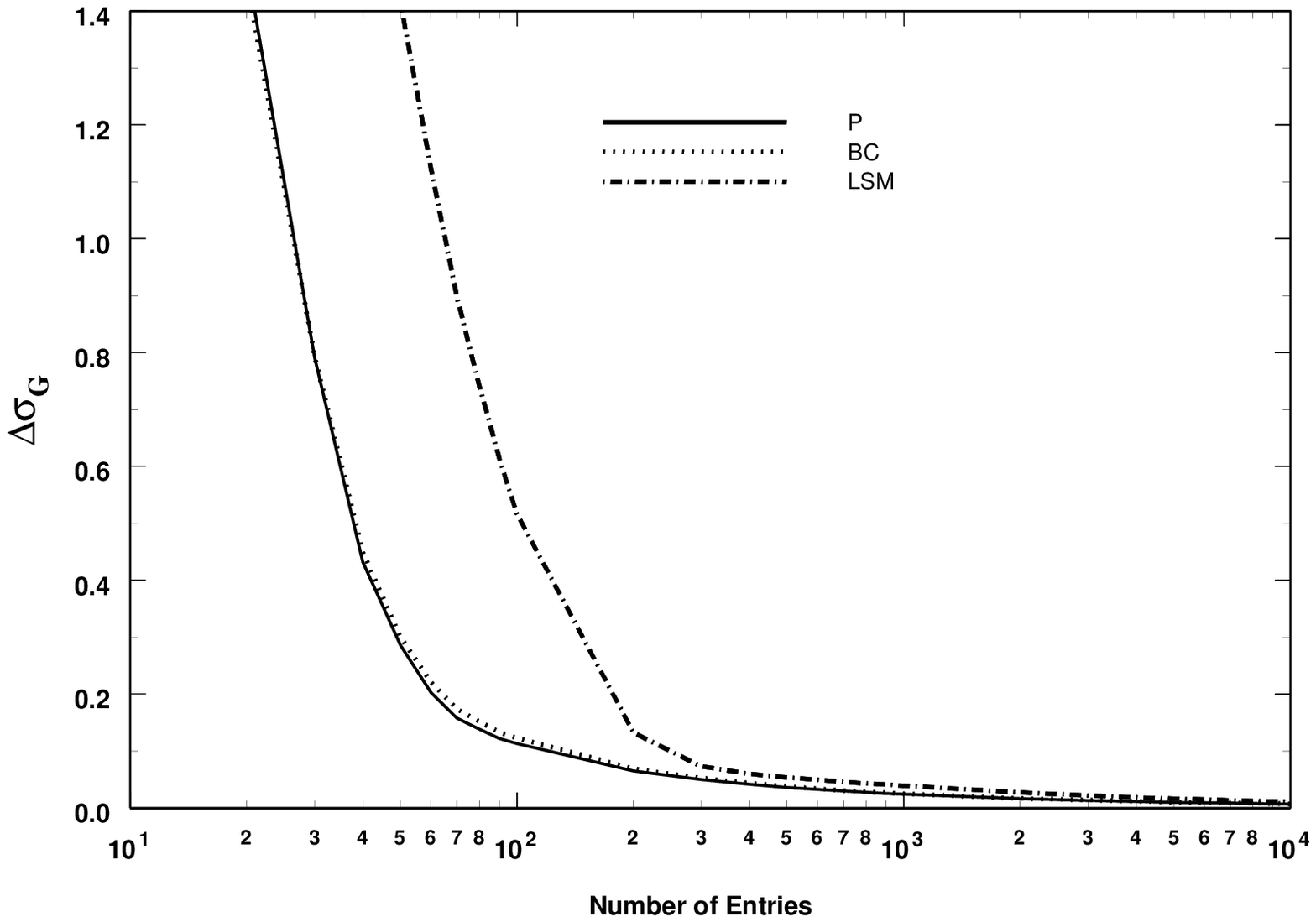}} 
\end{sideways}
\end{center}
\caption{ }
\end{figure}

\begin{figure} 
\begin{center}
\begin{sideways}
\mbox{\epsfig{file=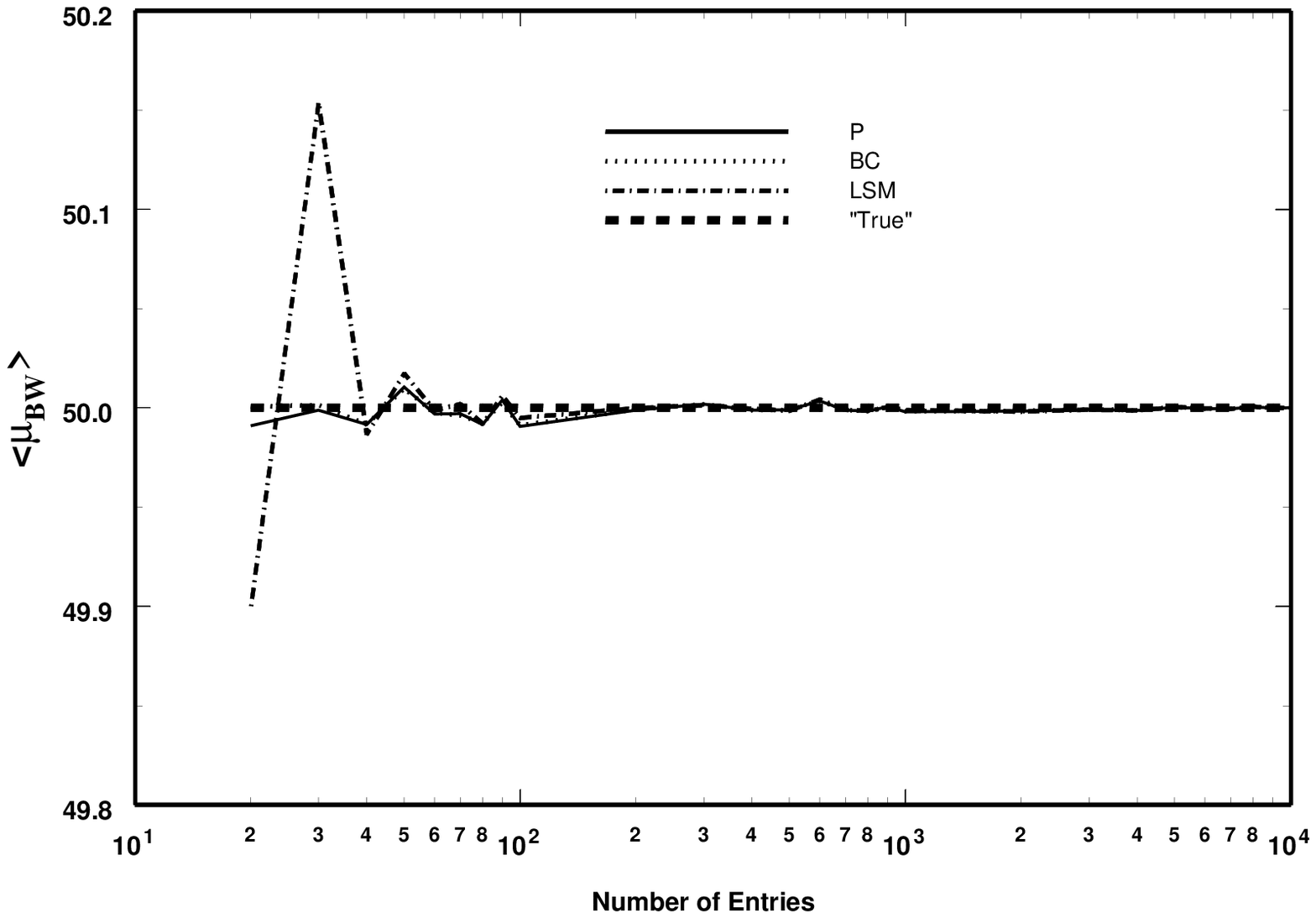}} 
\end{sideways}
\end{center}
\caption{ }
\end{figure}

\begin{figure} 
\begin{center}
\begin{sideways}
\mbox{\epsfig{file=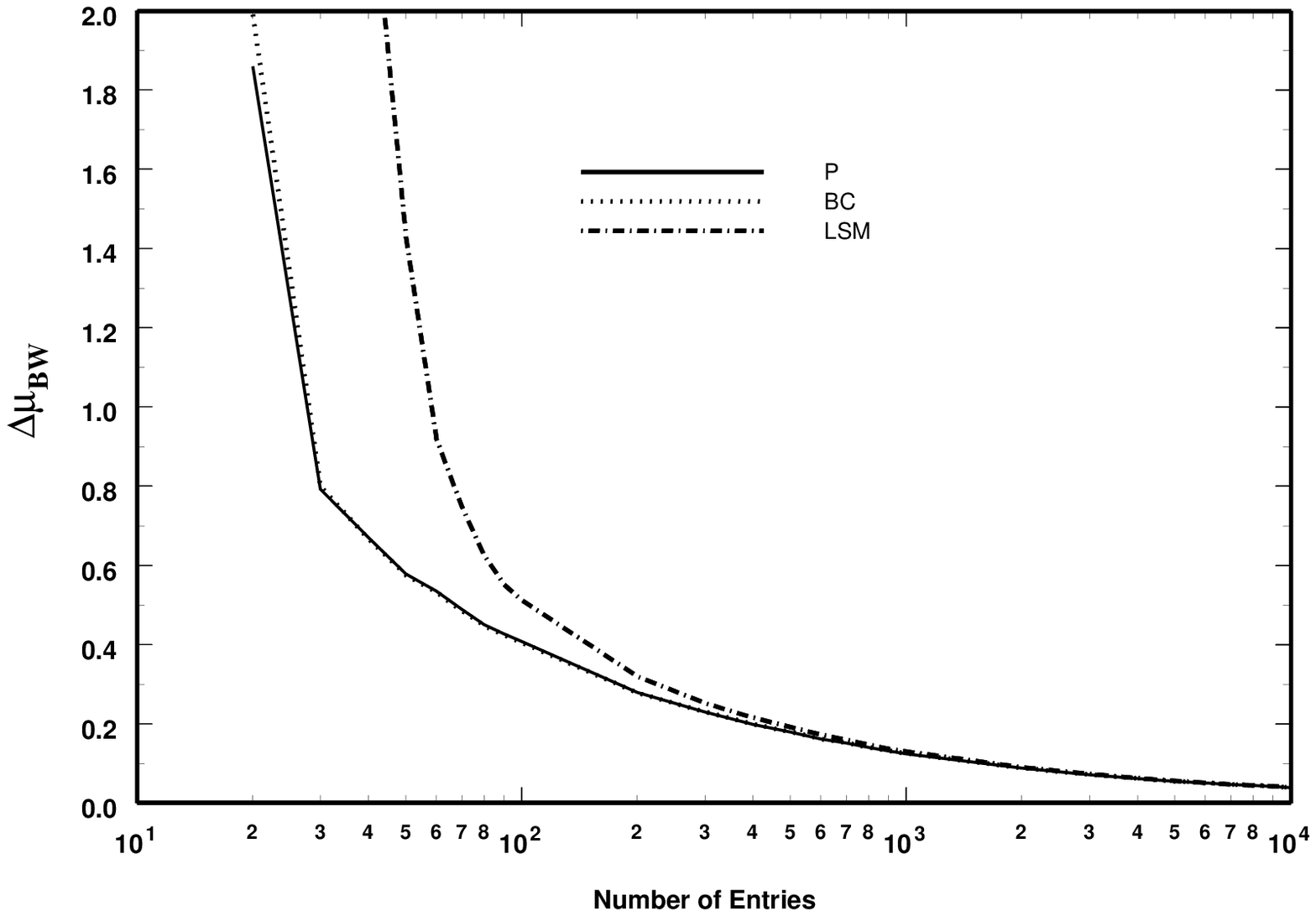}} 
\end{sideways}
\end{center}
\caption{ }
\end{figure}

\begin{figure} 
\begin{center}
\begin{sideways}
\mbox{\epsfig{file=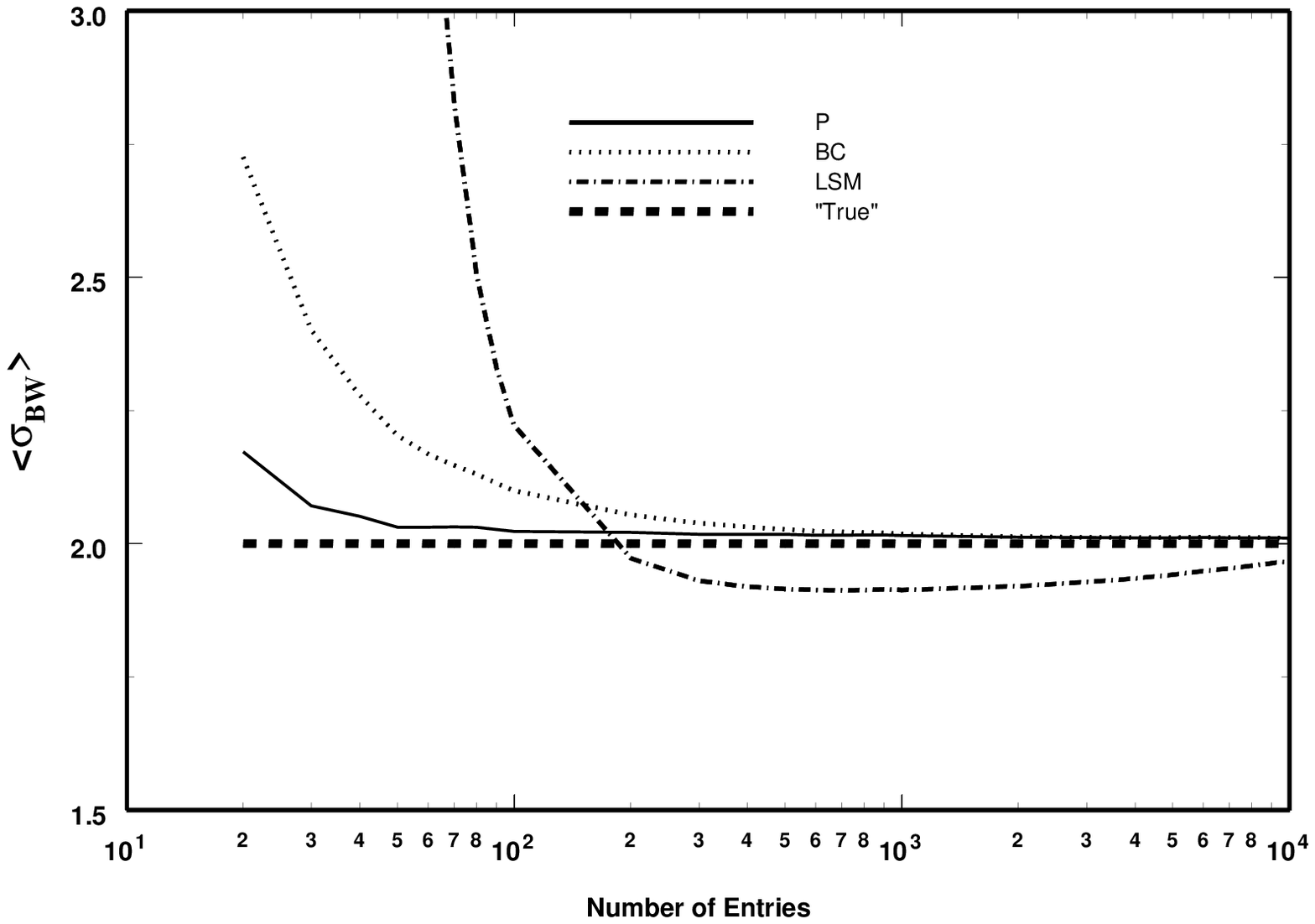}} 
\end{sideways}
\end{center}
\caption{ }
\end{figure}

\begin{figure} 
\begin{center}
\begin{sideways}
\mbox{\epsfig{file=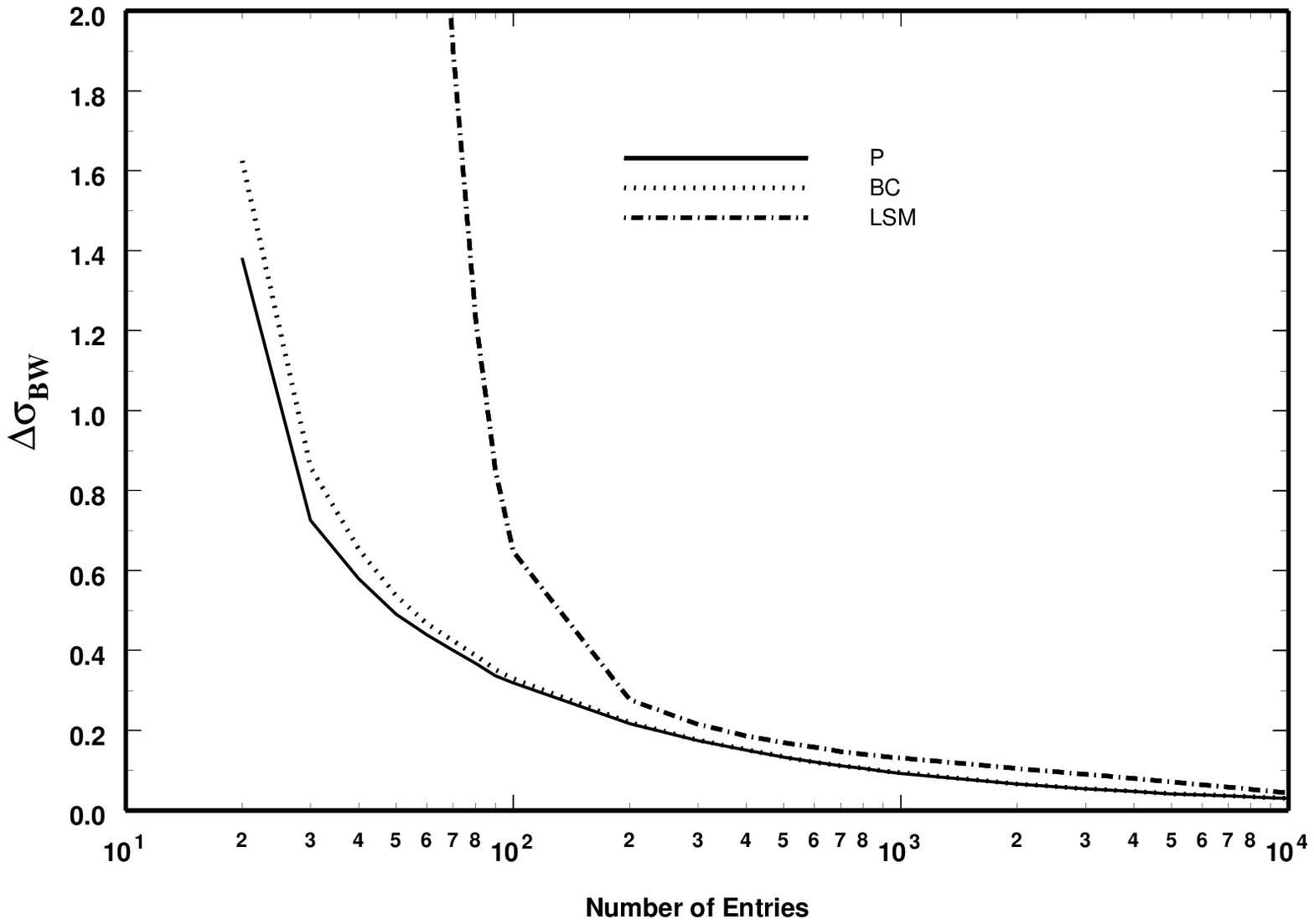}} 
\end{sideways}
\end{center}
\caption{ }
\end{figure}

\begin{figure} 
\begin{center}
\begin{sideways}
\mbox{\epsfig{file=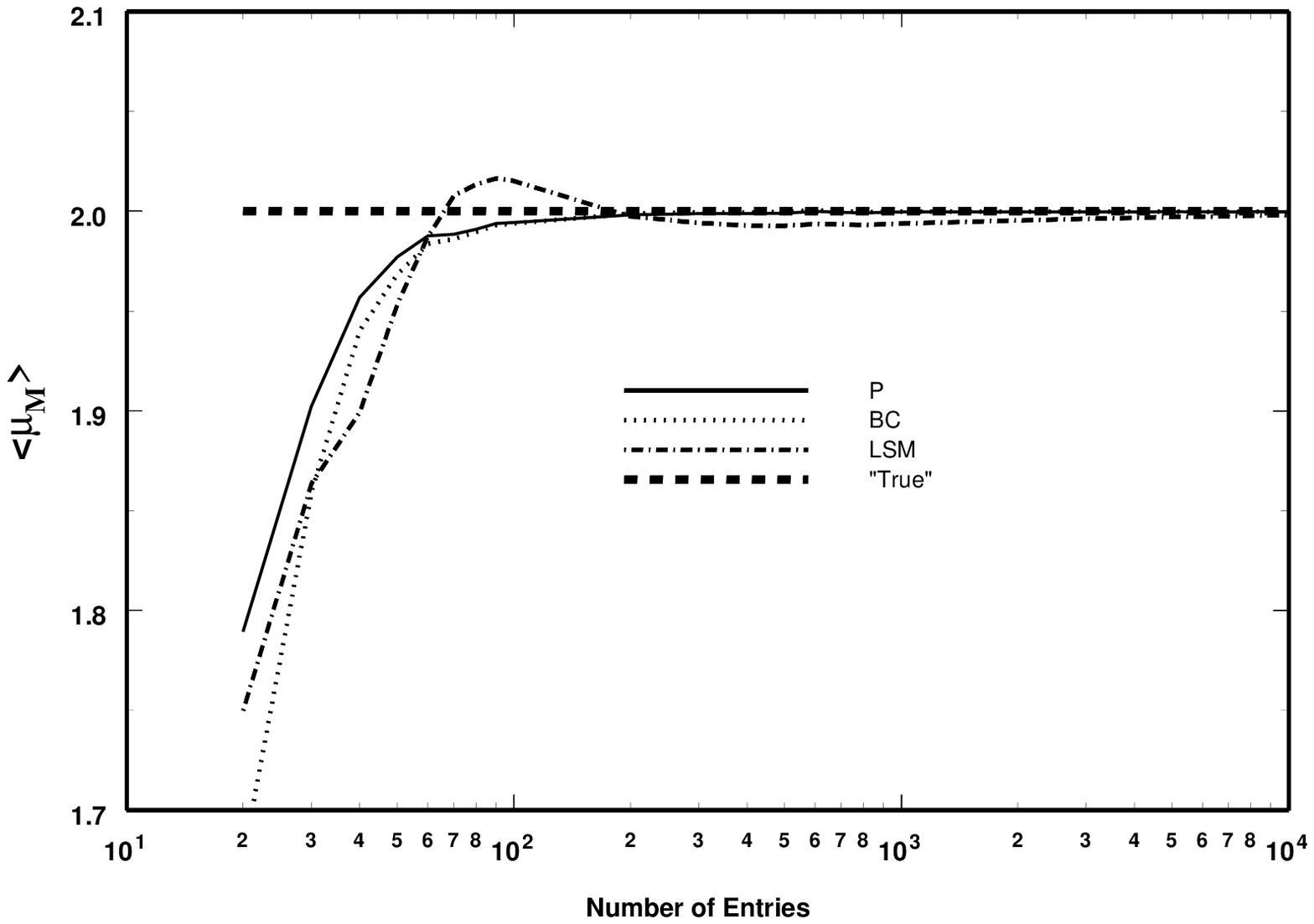}} 
\end{sideways}
\end{center}
\caption{ }
\end{figure}

\begin{figure} 
\begin{center}
\begin{sideways}
\mbox{\epsfig{file=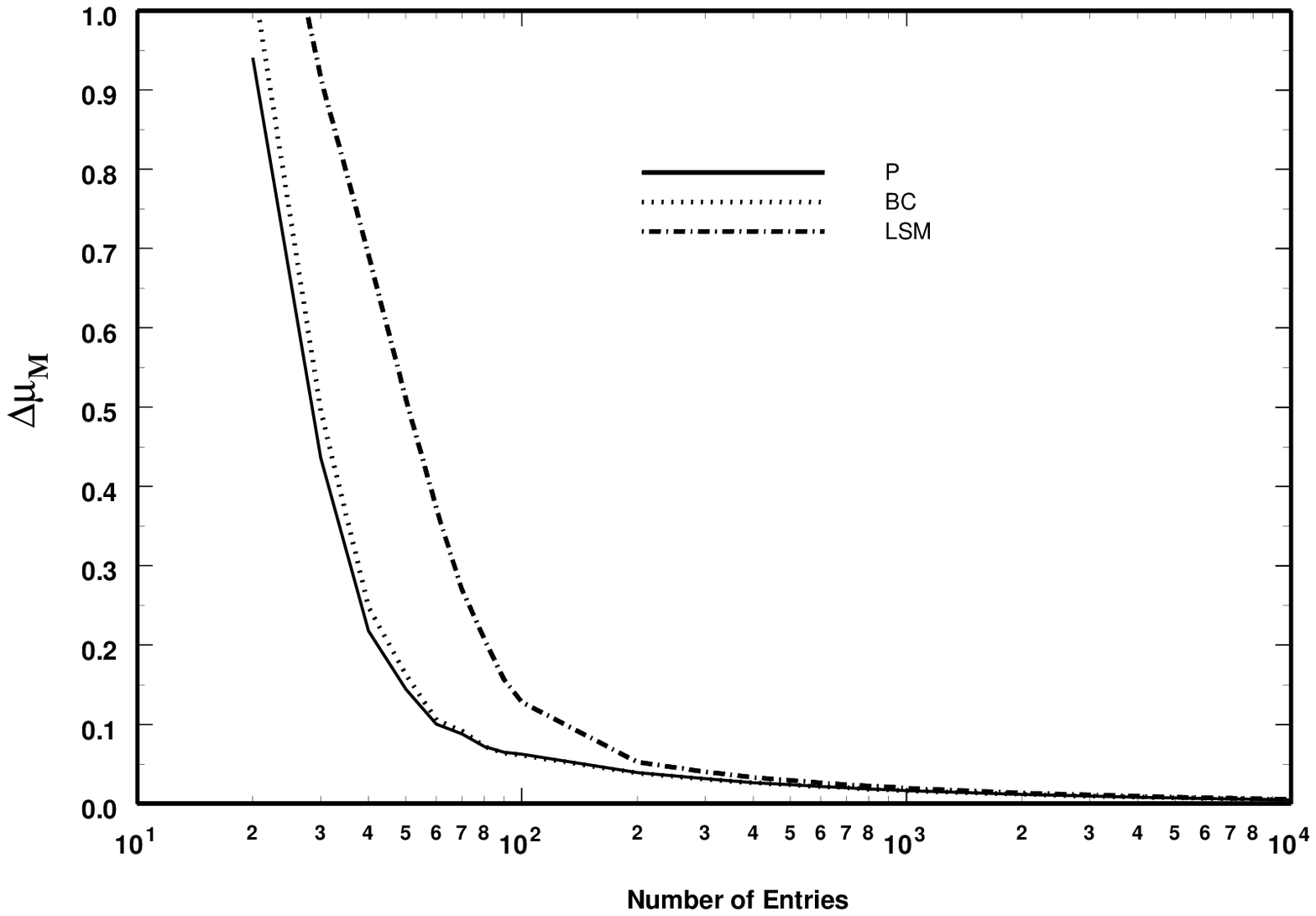}} 
\end{sideways}
\end{center}
\caption{ }
\end{figure}

\begin{figure} 
\begin{center}
\begin{sideways}
\mbox{\epsfig{file=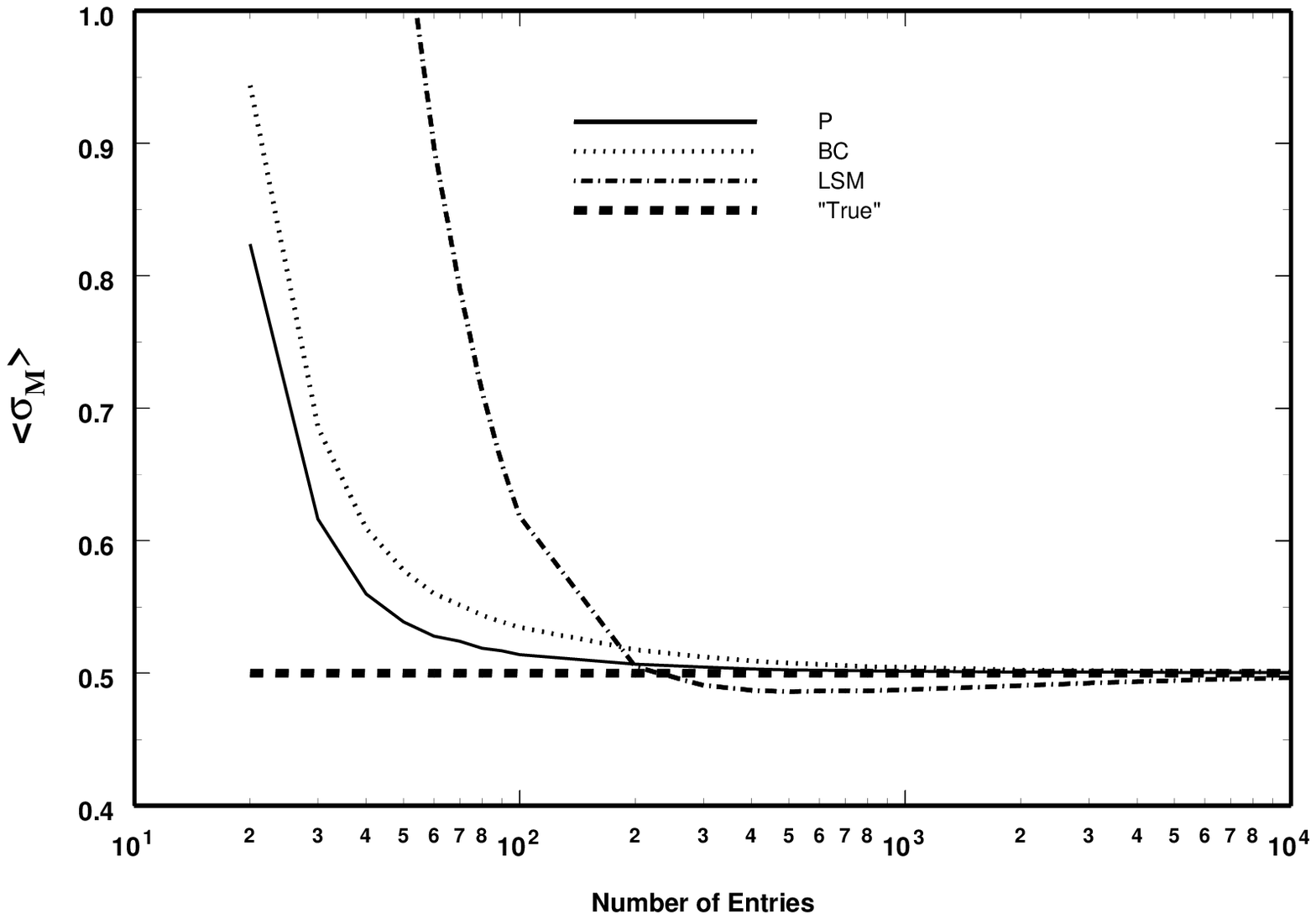}} 
\end{sideways}
\end{center}
\caption{ }
\end{figure}

\begin{figure} 
\begin{center}
\begin{sideways}
\mbox{\epsfig{file=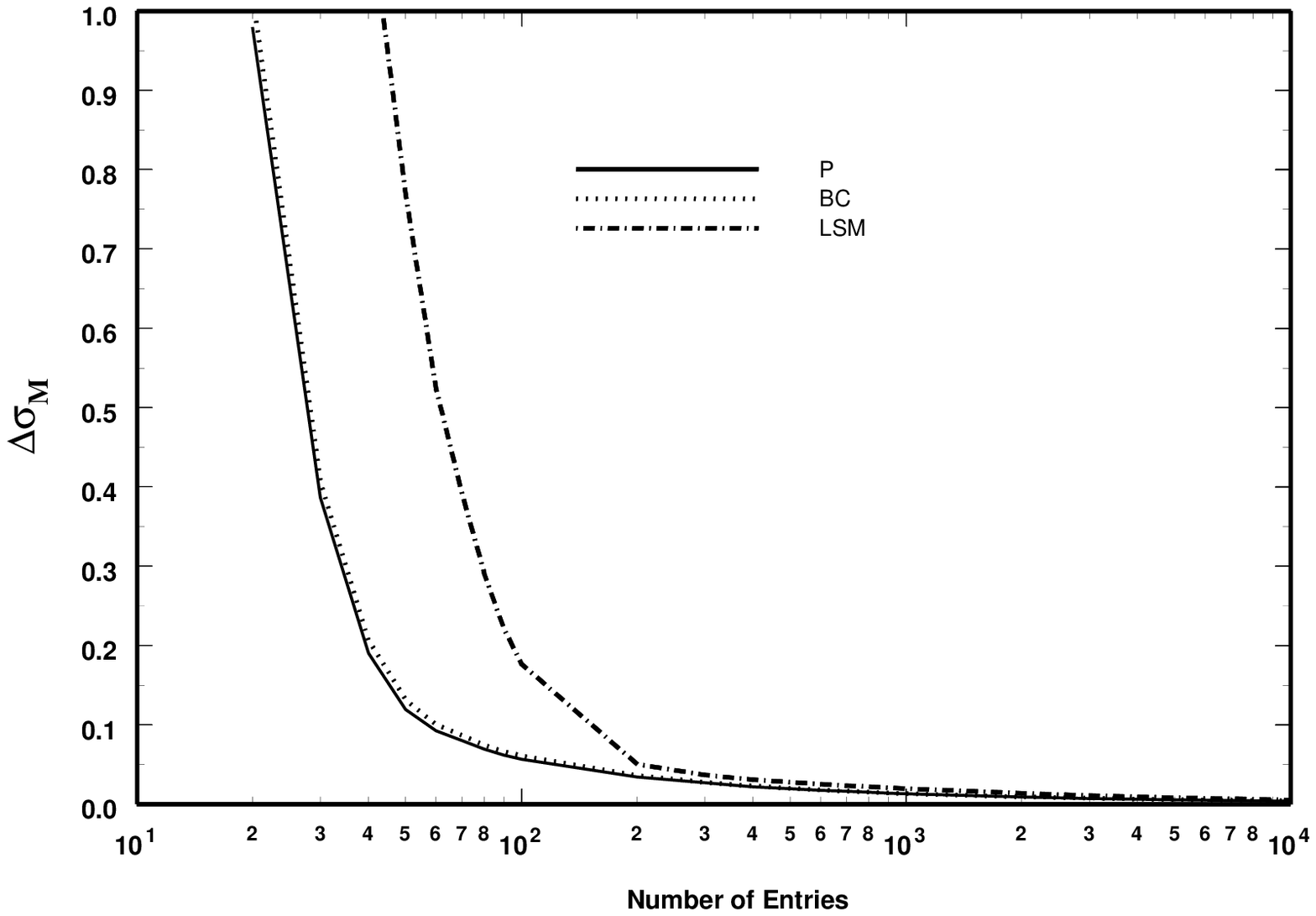}} 
\end{sideways}
\end{center}
\caption{ }
\end{figure}

\end{document}